\documentclass[10pt,english,twocolumn, conference]{IEEEtran}

\usepackage[T1]{fontenc}
\usepackage[latin1]{inputenc}
\usepackage{geometry}
\usepackage{amsthm}
\usepackage{amsmath}
\usepackage{amssymb}
\usepackage{graphicx}
\usepackage{epstopdf}
\usepackage{cite}
\usepackage{color}
\usepackage{mathtools}
\usepackage{cases}
\usepackage{stackengine}
\usepackage{accents}

\newcommand\blfootnote[1]{%
  \begingroup
  \renewcommand\thefootnote{}\footnote{#1}%
  \addtocounter{footnote}{-1}%
  \endgroup
}

\newtheorem{thm}{Theorem}

\newtheorem{rem}{Remark}

\renewcommand\appendix{\par
\setcounter{section}{0}
\setcounter{subsection}{0}
\setcounter{figure}{0}
\setcounter{table}{0}
\renewcommand\thesection{ \Alph{section}}
\renewcommand\thefigure{\Alph{section}\arabic{figure}}
\renewcommand\thetable{\Alph{section}\arabic{table}}
}

\usepackage{babel}

\allowdisplaybreaks

\begin{document}

\title{SWIPT Signalling over Complex AWGN Channels with Two Nonlinear Energy Harvester Models}

\author{Morteza Varasteh, Borzoo Rassouli, Hamdi Joudeh and Bruno Clerckx\\
Department of Electrical and Electronic Engineering, Imperial College London, London, U.K.\\
\{m.varasteh12; b.rassouli12; hamdi.joudeh10; b.clerckx\}@imperial.ac.uk. }

\maketitle

\begin{abstract}
Simultaneous Wireless Information and Power Transfer (SWIPT) is subject to nonlinearity at the energy harvester that leads to significant changes to transmit signal designs compared to conventional wireless communications. In this paper, the capacity of a discrete time, memoryless and complex Additive White Gaussian Noise (AWGN) channel in the presence of a nonlinear energy harvester at the receiver is studied. Considering the two common nonlinear energy harvester models introduced in the literature, two sets of constraints are considered. First the capacity is studied under average power (AP), peak amplitude (PA) and receiver delivery power (RDP) constraints. The RDP constraint is modelled as a linear combination of even-moment statistics of the channel input being larger than a threshold. It is shown that the capacity of an AWGN channel under AP and RDP constraints is the same as the capacity of an AWGN channel under an AP constraint, however, depending on the two constraints, it can be either achieved or arbitrarily approached. It is also shown that under AP, PA and RDP constraints, the amplitude of the optimal inputs is discrete with a finite number of mass points. Next, the capacity is studied under AP, PA and output outage probability (OOP) constraints. OOP is modelled as satisfying a certain probability inequality for the amplitude of the received signal being outside of a given interval. Similarly, it is shown that the amplitude of the optimal input is discrete with a finite number of mass points.\blfootnote{This work has
been partially supported by the EPSRC of UK, under grant EP/P003885/1
}
\end{abstract}

\section{Introduction}\label{Sec_Intro}

%Radio-Frequency (RF) waves can be utilized for transmission of both information and power simultaneously. As one of the primary works in the information theory literature, Varshney studied this problem in \cite{Varshney_2008}, in which he characterized the capacity-power function for a point-to-point discrete memoryless channel (DMC). He showed the existence of a tradeoff between the information rate and the delivered power for some channels, such as, point-to-point binary channels and amplitude constraint Gaussian channels. Recent results in the literature have also revealed that in many scenarios, there is a tradeoff between information rate and delivered power. Just to name a few, frequency-selective channel \cite{Grover_Sahai_2010}, MIMO broadcasting \cite{Zhang_Keong_2013}, interference channel \cite{Park_Clerckx_2013}.

One of the major efforts in a Simultaneous Wireless Information and Power Transfer (SWIPT) architecture is to increase the Direct-Current (DC) power and information rate at the output of the energy harvester and information decoder, respectively, without increasing the transmit power. As one of the primary works in the information theory literature, Varshney studied this problem in \cite{Varshney_2008}, in which he characterized the capacity-power function for a point-to-point discrete memoryless channel. The harvester, known as rectenna, is composed of an antenna followed by a rectifier.\footnote{In the literature, the rectifier is usually considered as a nonlinear device (usually a diode) followed by a low-pass filter. The diode is the main source of nonlinearity induced in the system.} In \cite{Trotter_Griffin_Durgin_2009,Clerckx_Bayguzina_2016}, it is shown that as a consequence of the rectifier nonlinearity, the RF-to-DC conversion efficiency is a function of rectenna's structure, as well as its input waveform. Accordingly, in order to maximize rectenna's DC power output, a systematic waveform design is crucial to make the best use of an available RF spectrum \cite{Clerckx_Bayguzina_2016}. Following \cite{Clerckx_Bayguzina_2016}, it was shown in \cite{Clerckx_2016,Varasteh_Rassouli_Clerckx_ITW} that the choice of a suitable input distribution (and therefore modulation and waveform) for SWIPT is affected by the rectifier nonlinearity and motivates the study of the capacity of AWGN channels under nonlinear power constraints.

In the literature, so far, depending on the application and available resources, two different models of the energy harvester nonlinearity are used. The first model is based on Taylor expansion of the diode characteristic function and is introduced in \cite{Clerckx_Bayguzina_2016}. It is shown that the harvested power is a function of summation of even moments of the received RF signal, which can be approximated with an acceptable level of accuracy by truncating it to the second and fourth moments. This model is appropriate for practical low power applications (with input power to the energy harvester of the order of $0$dBm or less). The other model, which is a function of the RF power of the received signal, is introduced in \cite{Boshkovska_NG_Zlatanov_Schober}. In this model, the rectenna's input/output power relationship is expressed in terms of a logistic (sigmoidal) function. This model is appropriate in applications where there is little guarantee for the energy harvester to operate below the diode breakdown edge.

The capacity of deterministic, complex and real, discrete-time memoryless AWGN channels has been investigated in the literature under various constraints, extensively. It seems that the linear AWGN channel subject to transmit average power constraint is an exception where the optimal input is Gaussian distributed \cite{Shannon_1948}, and under many other constraints, the optimal input leads to discrete inputs. To mention a few, \cite{Smith:IC:71} for a real AWGN channel with average power and amplitude constrained inputs, \cite{Shamai_BarDavid_1995} for complex AWGN channels with average and peak-power constraint and \cite{AbouFaycal_Trott_Shamai_2001} for complex Rayleigh-fading channel when no channel state information (CSI) is assumed either at the receiver or the transmitter.

Motivated by the two nonlinear models in the literature, we provide a step closer at identifying the fundamental limits of SWIPT structures. In this paper, we study a complex and discrete time memoryless AWGN channel under two sets of constraints. First, in order to tackle the nonlinear model introduced in \cite{Clerckx_Bayguzina_2016}, we study the capacity problem under average power (AP), peak amplitude (PA) and receiver delivery power (RDP) constraints. Next, to tackle the model in \cite{Boshkovska_NG_Zlatanov_Schober}, we study the capacity problem under average power (AP), peak amplitude (PA) and output outage probability (OOP) constraints. First, focusing on the nonlinear model of the harvester in \cite{Clerckx_Bayguzina_2016}, we show that the capacity of an AWGN channel under AP and RDP constraints is the same as the capacity of an AWGN channel under AP constraint. However, depending on the two constraints, the capacity can be either achieved or approached arbitrarily (irrespectively of the RDP constraint). We show that in line with the results reported in \cite{Smith:IC:71,Shamai_BarDavid_1995,AbouFaycal_Trott_Shamai_2001,Fahs_AbouFaycal_2012} the optimal input distributions are discrete with a finite number of mass points. The system model studied in this paper focuses on the nonlinearities at the receiver (over complex AWGN channels) and  indeed can be considered as a reciprocal of \cite{Fahs_AbouFaycal_2012}, where the main focus is on the nonlinearities at the transmitter (transmit nonlinear constraints as well as nonlinear channel inputs over real AWGN channels). Second, focusing on the nonlinear model of the harvester in \cite{Boshkovska_NG_Zlatanov_Schober}, we study the capacity problem under AP, PA and OOP constraint, where similar results (discrete inputs with a finite number of mass points) are obtained.

\textit{Organization}: In Section \ref{Sec_System_Model}, we introduce the system model and define the channel capacity problem studied here. In Section \ref{Sec_Main_res}, we introduce the main results of the paper. In Section \ref{Sec_Num_res}, numerical results are illustrated. The proofs of the provided results are summarized due to space limit, and where it is necessary, we refer to the details in the extended version of the paper \cite{Varasteh_Rassouli_Clerckx_17}.

\textit{Notations}: Throughout this paper, the standard circular symmetric complex Gaussian (CSCG) distribution is denoted by $\mathcal{CN}(0,1)$. $F_{\pmb{r}}(r)$ and $f_{\pmb{r}}(r)$ denote, respectively, cumulative distribution function (CDF) and the probability density function (pdf) of thr random variable $\pmb{r}$, with its support denoted as $\text{supp}\{\pmb{r}\}$. We define the kernel $K(R,r)\triangleq R e^{-\frac{R^2+r^2}{2}} I_0(rR)$, where $I_0(x)=1/\pi\int_{0}^{\pi}e^{x\cos(\theta)}d\theta$ is the modified Bessel function of the first kind and order zero. The marcum Q-function is defined as $Q(r,A)\triangleq\int_{A}^{\infty}K(R,r)dR$. The error function is defined as $\text{erf}(x)=2/\sqrt{\pi}\int_{0}^{x}e^{-t^2}dt$.

\section{System Model, Problem Definition and Preliminaries}\label{Sec_System_Model}
Consider the following complex representation of a discrete-time AWGN channel,
\begin{align}\label{E69}
\pmb{y}_k=\pmb{x}_k+\pmb{n}_k,
\end{align}
where $\{\pmb{y}_k\}$, $\{\pmb{x}_k\}$ and $\{\pmb{n}_k\}$ represent the sequences of complex-valued samples of the channel output, input and AWGN, respectively, and $k$ is the discrete-time index. The real and imaginary parts of the signal $\{\pmb{y}_k\}$ indicate the Inphase and Quadrature components, respectively. The noise samples $\{\pmb{n}_k\}$ are assumed to be CSCG distributed as $\mathcal{CN}(0,2)$.\footnote{Since the channel (\ref{E69}) is stationary and memoryless, the capacity achieving statistics of the input are also memoryless, and hence with no loss of generality we suppress the time index throughout the paper.}

\subsection{Capacity under AP, PA and RDP constraints}
The nonlinear model for the rectenna introduced in \cite{Clerckx_Bayguzina_2016} is a function of even moments of the received Radio Frequency (RF) signal. In \cite{Varasteh_Rassouli_Clerckx_17}, it is shown that the modelled harvested power can be lower bounded by considering the even moments of the baseband equivalent of the RF signal. Motivated by that, the capacity of a discrete-time complex AWGN channel \cite[Ch. 7]{Gallager_book} under AP, PA and RDP is given by
\begin{equation}\label{E3}
\begin{aligned}
C_{P_a,P_d,r_p}\triangleq& \sup\limits_{p_{\pmb{x}}(x)}
& & I(\pmb{x};\pmb{y}) \\
& \text{s.t.}:
& & \left\{\begin{array}{lll}
\mathbb{E}[|\pmb{x}|^2]  \leq  P_a, \\
\mathbb{E}[g(|\pmb{x}|)]\geq P_d ,\\
|\pmb{x}|\leq r_p,
\end{array}\right.
\end{aligned}
\end{equation}
where $P_a$, $P_d$ and $r_p$ are the transmitter AP, PA and RDP constraints, respectively. $g(\cdot)$ is assumed to be a continuous positive function having the form of
\begin{align}\label{E54}
g(r)=\sum\limits_{i=0}^{n}\alpha_i r^{2i},~r\geq 0,
\end{align}
where $n\geq 2$ is an arbitrary integer. Note that since $g(r)$ is assumed to be a positive function, we have $\alpha_n>0$, and hence, $\lim_{r\rightarrow\infty}g(r)=\infty$.\footnote{The scenario $g(r)=\alpha_0+\alpha_2r^2$ beside AP constraint and beside AP/ PA constraints are straightforward, yielding that CSCG distributions \cite{Shannon_1948} and discrete distributions with a finite number of mass points are optimal \cite{Shamai_BarDavid_1995}, respectively. Accordingly, we will not consider it throughout the paper.}

\subsection{Capacity under AP, PA and OOP constraints}
The model for energy harvester proposed in \cite{Boshkovska_NG_Zlatanov_Schober}, captures both the threshold effect\footnote{This relates to the sensitivity of the harvester in the low-power regime and is also indirectly captured by/connected to the nonlinear model (\ref{E54}) and waveform design  \cite{Trotter_Griffin_Durgin_2009,Clerckx_Bayguzina_2016}.} (the power level below which the rectifier turns off and denoted by $A_l^2$) and the saturation effect (the power level above which the rectifier's power gain remains constant and denoted by $A_u^2$). From an energy harvesting point of view, it is most favourable to keep the energy harvester operating without getting saturated. Motivated by this nonlinear model, we introduce the OOP constraint as $\textrm{Pr}(|\pmb{x}|\geq A_u\cup|\pmb{x}|\leq A_l)\leq 1-\epsilon$ for $0\leq A_l<A_u\leq \infty$ and $0\leq \epsilon <1$. We note that since this criterion is equivalent to $\textrm{Pr}(A_l\leq |\pmb{y}|\leq A_u)\geq \epsilon$, in the following for simplicity we consider the latter as the OOP constraint. The capacity of a complex discrete-time AWGN channel \cite[Chapter 7]{Gallager_book} under AP, PA and OOP is given by
\begin{equation}\label{E1}
\begin{aligned}
C_{P_a,A_l,A_u,\epsilon,r_p}\triangleq& \sup\!
& &I(\pmb{x};\pmb{y}) \\
& p_{\pmb{x}}(x):
& & \left\{\begin{array}{lll}
\mathbb{E}[|\pmb{x}|^2]  \leq  P_a,\\
|\pmb{x}|< r_p, \\
\textrm{Pr}(A_l\leq |\pmb{y}|\leq A_u)\geq \epsilon,
\end{array}\right.
\end{aligned}
\end{equation}

\subsection{Preliminaries}
Expressing $I(\pmb{x};\pmb{y})$ in terms of differential entropies, we have
\begin{align}\label{E2}
I(\pmb{x};\pmb{y})&=h(\pmb{y})-\log 2\pi e.
\end{align}
Therefore, the problems in (\ref{E3}) and (\ref{E1}) are equivalent to the supremization of differential entropy $h(\pmb{y})$ subject to the corresponding constraints in (\ref{E3}) and (\ref{E1}), respectively. Using polar coordinates\footnote{The polar representation simplifies the problem, since the constraints are circular.} $\pmb{x}=\pmb{r}e^{i\pmb{\theta}}$ and $\pmb{y}=\pmb{R}e^{i\pmb{\phi}}$ ($\pmb{r},\pmb{R}\geq 0$ and $\pmb{\theta},\pmb{\phi}\in[0,2\pi)$) and following the same steps in \cite{Shamai_BarDavid_1995}, we have
\begin{align}\label{E4}
h(\pmb{y})\leq -\int\limits_{0}^{\infty} f_{\pmb{R}}(R;F_{\pmb{r}})\ln \frac{f_{\pmb{R}}(R;F_{\pmb{r}})}{R} dR +\ln 2\pi,
\end{align}
where $f_{\pmb{R}}(R;F_{\pmb{r}})$ is the pdf of $\pmb{R}$ induced by $F_{\pmb{r}}$ given as $f_{\pmb{R}}(R;F_{\pmb{r}})=\int_{0}^{r_p}K(R,r)dF_{\pmb{r}}(r)$. Note that by selecting $\pmb{r}$ and $\pmb{\theta}$ independent with uniformly distributed $\pmb{\theta}$ over $[0,2\pi)$, the upperbound in (\ref{E4}) becomes tight (for more details see \cite{Shamai_BarDavid_1995}) and we have $f_{\pmb{R},\pmb{\phi}}(R,\phi)=\frac{1}{2\pi}f_{\pmb{R}}(R;F_{\pmb{r}})$.

In (\ref{E1}), the OOP constraint can be equivalently written as
\begin{align}
  \textrm{Pr}(A_l\leq |\pmb{y}|\leq A_u)&=\int\limits_{0}^{\infty} \left(Q(r,A_l)-Q(r,A_u) \right)dF_{\pmb{r}}(r).
\end{align}

Therefore, the optimization problem in (\ref{E3}) and (\ref{E1}) are reduced to the following problems, respectively
\begin{align}\label{E19}
C_{P_a,P_d,r_p}=& \underset{F_{\pmb{r}}\in \Omega_1\cap\Omega_2}\sup H(F_{\pmb{r}}), \\\label{E5}
C_{P_a,A_l,A_u,\epsilon,r_p}=& \underset{F_{\pmb{r}}\in \Omega_1\cap\Omega_3}\sup H(F_{\pmb{r}}),
\end{align}
where $F_{\pmb{r}}$ is such that $F_{\pmb{r}}(0^{-})=0,F_{\pmb{r}}(r_p)=1$ and $H(F_{\pmb{r}})$, $\Omega_1$, $\Omega_2$ and $\Omega_3$ are given as
\begin{align}\label{E8}
H(F_{\pmb{r}})&\triangleq -\int\limits_{0}^{\infty} f_{\pmb{R}}(R;F_{\pmb{r}})\ln \frac{f_{\pmb{R}}(R;F_{\pmb{r}})}{R} dR,
\end{align}
\vspace{-5mm}
\begin{subequations}\label{E47}
\begin{align}\label{E47a}
  \Omega_1&\!=\!\left\{\!\!F_{\pmb{r}}\!:\!\!\int_{0}^{r_p}r^2dF_{\pmb{r}}(r)\!\leq\! P_a\right\},\\\label{E47b}
  \Omega_2&\!=\!\left\{\!\!F_{\pmb{r}}\!:\!\!\int_{0}^{r_p}g(r)dF_{\pmb{r}}(r)\!\geq\! P_d\right\},\\
  \Omega_3&\!=\!\left\{\!\!F_{\pmb{r}}\!:\!\!\int_{0}^{r_p}\left(Q(r,A_l)\!-\!Q(r,A_u)\right)dF_{\pmb{r}}(r)\!\geq\! \epsilon\right\}.
\end{align}
\end{subequations}

\section{Main results}\label{Sec_Main_res}
In this section, we provide the main results of this paper regarding the problems introduced in (\ref{E19}) and (\ref{E5}).

\subsection{Capacity under AP, PA and RDP constraints}
In the following, we first characterize the capacity in (\ref{E19}) when the channel input amplitude constraint is $r_p=\infty$. In the next theorem, we study the capacity problem in (\ref{E19}) when $r_p<\infty$. We accordingly, derive the necessary and sufficient condition for the optimal distributions achieving the capacity.

\begin{thm}\label{T0}
The capacity of the channel in (\ref{E69}) under AP, RDP constraints and $r_p=\infty$, i.e., $C_{P_a,P_d,\infty}$ is characterized as
\begin{align}
 C_{P_a,P_d,\infty} =\log(1+P_a/2).
\end{align}
If $P_d\leq P_R$, the capacity $C_{P_a,P_d,\infty}$ is achieved by a unique input distributed as $\pmb{x}\sim\mathcal{CN}(0,P_a)$, and for $P_d> P_R$, the capacity $C_{P_a,P_d,\infty}$ is not achieved, however, can be approached arbitrarily, where $P_R=\frac{1}{P_a}\int_{0}^{\infty}rg(r)e^{-\frac{r^2}{2P_a}}dr$ is the RDP corresponding to CSCG input.
\end{thm}
\textit{Proof}: It can be verified that the space $\Omega_1\cap\Omega_2$ is convex. For $r_p=\infty$, the space $\Omega_1\cap\Omega_2$ is not compact (see \cite[Lem. 3]{Varasteh_Rassouli_Clerckx_17} for a counterexample contradicting the compactness). It is shown in \cite[Lem. 10]{Varasteh_Rassouli_Clerckx_17} that for every $F_{\pmb{r}}\in \Omega_1\cap\Omega_2$, $H(F_{\pmb{r}})$ exists, and is continuous, concave and weakly differentiable. Strict concavity of $H(F_{\pmb{r}})$ over $\Omega_1\cap\Omega_2$ is proved by noting that the integral transform $f_{\pmb{R}}(R;F_{\pmb{r}})=\int_{0}^{\infty}K(R,r)dF_{\pmb{r}}(r)$ is invertible (for the proof see \cite[App. II]{Shamai_BarDavid_1995}). For $P_d\leq P_R$, the RDP constraint is inactive and the optimal input is $\pmb{x}\sim\mathcal{CN}(0,P_a)$. For $P_d> P_R$, it is easily verified that for a fixed $P_a$, the capacity $C_{P_a,P_d,\infty}$ is non-increasing with $P_d$. Therefore, we have $C_{P_a,P_R,\infty}\geq C_{P_a,P_d,\infty}$ for $P_d>P_R$. Also due to the strict concavity of $H(F_{\pmb{r}})$, the capacity achieving distributions are unique. Accordingly, we have $C_{P_a,P_R,\infty}> C_{P_a,P_d,\infty}$. In the following, we show that there exist distributions that can approach the $C_{P_a,P_R,\infty}$ arbitrarily.

 Consider the following sequence of distribution functions
\begin{align}
F_{\pmb{r}_l}(r)=\left\{\begin{array}{ll}
                 0&r<0\\
                 1-\frac{1}{l^2}   & 0\leq r<\sqrt{P_a}l \\
                 1 & r\geq  \sqrt{P_a}l
               \end{array}\right.,~l=2,\ldots.
\end{align}
Note that $\mathbb{E}_{F_{\pmb{r}_l}}[\pmb{r}_l^2]=P_a$, hence, satisfying the AP constraint. Also, for the RDP constraint we have
\begin{align}
P_{d,l}\triangleq\mathbb{E}_{F_{\pmb{r}_l}}[g(\pmb{r}_l)]=\alpha_0+ \alpha_1 P_a + \sum\limits_{i=2}^{n}\alpha_i P_a^i l^{2i-2}.
\end{align}
Since $n\geq 2$ by construction, it is guaranteed that there exists an integer number $L$, such that for $l>L$, $P_{d,l}\geq P_d$ (note that $P_{d,l}$ increases with $l$). Due to convexity of the optimization space, time sharing is valid in our system model. Hence, we can construct a complex input $\pmb{x}_l$ with its phase $\pmb{\theta}_l$ uniformly distributed over $[0,2\pi)$ and its amplitude $\pmb{r}_l$ distributed according to the following CDF
\begin{align}\label{E80}
F_{\pmb{r}_{ts}}(r)\!=\!(1-\tau) F_{\pmb{r}_R}(r)+\tau F_{\pmb{r}_l}(r),\tau\in(0,1),l\!>\!L,
\end{align}
where the subscript $ts$ in $F_{\pmb{r}_{ts}}$ stands for time-sharing and $F_{\pmb{r}_R}(r)$ corresponds to the CDF of the Rayleigh distribution (recall that amplitude of a CSCG distributed complex random variable has a Rayleigh CDF). By choosing $\tau=(P_d-P_R)/(P_{d,l}-P_R)$, it is easy to verify that $0<\tau<1$ and  the constraints
\begin{align}
\mathbb{E}_{F_{\pmb{r}_{ts}}}[\pmb{r}_{ts}^2]\leq P_a,~ \mathbb{E}_{F_{\pmb{r}_{ts}}}[g(\pmb{r}_{ts})]\geq P_d,
\end{align}
are both satisfied. On the other hand, due to strict concavity of the entropy $H(F_{\pmb{r}})$, we have
\begin{align}\label{E79}
H(F_{\pmb{r}_{ts}})\!>\! (1-\tau) H(F_{\pmb{r}_R})\!+\!\tau H(F_{\pmb{r}_l}),\tau\in(0,1),l\!>\!L.
\end{align}
For a given $P_d$, we can increase $l$ arbitrarily. Since $P_{d,l}$ increases with $l$, therefore $\tau$ can be made arbitrarily close to zero. Rewriting (\ref{E79}), we have
\begin{align}
 H(F_{\pmb{r}_R})>H(F_{\pmb{r}_{ts}})> (1-\tau) H(F_{\pmb{r}_R})+\tau H(F_{\pmb{r}_l}),
\end{align}
where by letting $\tau$ tend to zero (equivalently letting $P_{d,l}\rightarrow\infty$) the result of Theorem \ref{T0} is concluded.
\qed

From the result of Theorem \ref{T0}, it is verified that for $n\geq 2$ in (\ref{E54}), the capacity of an AWGN channel in (\ref{E69}) for $r_p=\infty$ is independent of the value of the RDP constraint, i.e., $P_d$. That is, given $P_a$, the capacity $C_{P_a,P_d,\infty}$ is unchanged with $P_d$.

\begin{thm}\label{T1}
The optimal distribution denoted by $F_{\pmb{r}_1^{o}}$ achieving the capacity $C_{P_a,P_d,r_p}$ for $r_p<\infty$, is unique and the corresponding set of points of increase is finite (,i.e., the cardinality of the random variable $\pmb{r}_1^{o}$ is finite). Furthermore, $F_{\pmb{r}_1^{o}}$ is optimal if and only if there are unique parameters $\mu_1,\mu_2\geq 0$ for which
\begin{subequations}\label{E70}
\begin{align}
&h(r;F_{\pmb{r}_1^{o}})-\mu_1 r^2+\mu_2 g(r) -K_1=0,\forall r\!\in\!\emph{supp}\{\pmb{r}_1^{o}\},\\
&h(r;F_{\pmb{r}_1^{o}})-\mu_1 r^2+\mu_2 g(r) -K_1\leq 0,\forall r\!\in\! [0,r_p],
\end{align}
\end{subequations}
where $K_1\triangleq H(F_{\pmb{r}_1^o})-\mu_1 P_a+\mu_2 P_d$ and
\begin{align}\label{E14}
h(r;F_{\pmb{r}})&=-\int\limits_{0}^{\infty} K(R,r) \log{\frac{f(R,F_{\pmb{r}})}{R}}dR.
\end{align}
\end{thm}
\textit{Proof}: For $r_p<\infty$, the probability optimization space $\Omega_1\cap\Omega_2$ is convex and compact in the Levi metric \cite{Smith:IC:71}. This along with the existence, continuity, strict concavity and weak differentiability of $H(F_{\pmb{r}})$ over $F_{\pmb{r}}\in \Omega_1\cap\Omega_2$ (see \cite[Lem. 10]{Varasteh_Rassouli_Clerckx_17}) qualifies applying the Lagrangian theorem \cite[Sec. 7.4]{Luenberger_1969}. Hence, the necessary and sufficient condition for an input $\pmb{r}_1^0$ to be optimal is
\begin{align}\label{E6}
\!\!\int\limits_{0}^{r_p}\!\!(h(r;F_{\pmb{r}_1^o})\!-\!\mu_1 r^2\! \!+\!\mu_2 g(r) )dF_{\pmb{r}}(r)\!\leq\! K_1,\!\forall F_{\pmb{r}}\!\in \!\Omega_1\!\cap\!\Omega_2.
\end{align}
Noting that $K(R,r)$ is an analytic function (with respect to both of its arguments) and following a similar approach as in \cite{Smith:IC:71}, it can be verified that the amplitude of the optimal input $\pmb{r}_1^0$ is discrete with a finite number of mass points (see \cite{Varasteh_Rassouli_Clerckx_17} for more details). Note that for the optimal complex input $\pmb{x}$, it is enough to consider the phase $\pmb{\theta}$ uniformly distributed over $[0,2\pi)$.
\qed

\begin{rem}\label{Rem1}
By expanding $h(r;F_{\pmb{r}})$ in (\ref{E14}) (see \cite[App. C]{Varasteh_Rassouli_Clerckx_17} for more details), we have
\begin{align}\nonumber
&\int\limits_{0}^{\infty} K(R,r) \log{\frac{R}{f(R,F_{\pmb{r}})}}dR>-2.
\end{align}
Rewriting the KKT condition, for the inequality condition in (\ref{E70}) we get
\begin{align}\label{E9}
0\leq \mu_2\leq \frac{K_1+2+\mu_1 r^2}{g(r)},~r\in [0,r_p].
\end{align}
Since by definition, the function $g(r)$ grows faster than $r^2$, and (\ref{E9}) is valid for any $r\in[0,r_p]$, by substituting $r=r_p$, we have $\mu_2\rightarrow 0$ as $r_p\rightarrow \infty$. The intuition behind this is as follows. $\mu_2$ can be considered as the opposite sign of $\partial C_{P_a,P_d,r_p}/\partial r_p$. As $r_p$ increases, $C_{P_a,P_d,r_p}$ approaches $C_{P_a,P_d,\infty}$. From Theorem \ref{T0}, we already know that capacity $C_{P_a,P_d,\infty}$ is unchanged for any $P_d< \infty$. Therefore, $\partial C_{P_a,P_d,r_p}/\partial r_p\rightarrow0$ as $r_p\rightarrow\infty$.
\end{rem}

\subsection{Capacity under AP, PA and OOP constraints}
In the following, we derive the necessary and sufficient condition for the optimal distributions achieving the capacity in (\ref{E5}), when the input amplitude constraint is $r_p<\infty$.

\begin{thm}\label{T1}
The optimal distribution denoted by $F_{\pmb{r}_2^{o}}$ achieving the capacity $C_{P_a,A_l,A_u,\epsilon,r_p}$ for $r_p<\infty$, is unique and the corresponding set of points of increase is finite (,i.e., the cardinality of the random variable $\pmb{r}_2^{o}$ is finite). Furthermore, $F_{\pmb{r}_2^{o}}$ is optimal if and only if there are unique parameters $\lambda_1,\lambda_2\geq 0$ for which
\begin{subequations}\label{E7}
\begin{align}
&h(r;\!F_{\pmb{r}_2^{o}})\!-\!\lambda_1 r^2\!+\!\lambda_2 (\!Q(r,A_l) \!-\! Q(r,A_u)\!)\!=\!K_2,\!\forall r\!\in\! \emph{supp}\{\pmb{r}_2^{o}\},\\
&h(r;\!F_{\pmb{r}_2^{o}})\!-\!\lambda_1 r^2\!+\!\lambda_2 (\!Q(r,A_l)\! -\! Q(r,A_u)\!)\!\leq\!K_2,\!\forall r\!\in\! [0,r_p],
\end{align}
\end{subequations}
where $K_2\triangleq H(F_{\pmb{r}_2^o})-\lambda_1 P_a+\lambda_2 \epsilon$.
\end{thm}
\textit{Proof}: Noting that $Q(r,A_l) - Q(r,A_u)$ is a bounded and continuous function of $r$, the convexity and compactness of the optimization probability space $\Omega_1\cap\Omega_3$ is verified in a similar approach as in \cite[Appendix A]{AbouFaycal_Trott_Shamai_2001}\footnote{We note that the convexity and compactness hold valid for $r_p\leq \infty$}.  Therefore, by applying the Lagrangian theorem and following the same argument as in \cite{Smith:IC:71}, it is shown that it is necessary and sufficient for the optimal input $\pmb{r}_2^o$ to satisfy (\ref{E7}). Due to the fact that the function $Q(z,A_l) - Q(z,A_u)$ is analytic over $z$, using identity theorem and the fact that the distribution of amplitude of the received signal is continuous for every $F_{\pmb{r}}\in\Omega_1\cap\Omega_3$, it is shown that the optimal input does not contain a limit point in its support. This along with the fact that the support of the input is finite $r_p<\infty$, yields the desired result.
\qed

Note that the results in (\ref{E70}) and (\ref{E7}) are important in the sense that they can be utilized to obtain the optimal distributions using numerical programming.

\section{Numerical Results}\label{Sec_Num_res}
In this section, we provide some numerical illustrations of the results obtained in Section \ref{Sec_Main_res}. In the following, we first summarize the steps in obtaining the optimal inputs, and next, we illustrate the obtained numerical results.

To obtain the optimal inputs corresponding to the problems in (\ref{E3}) and (\ref{E1}), we resort to numerical programming. Accordingly, we solve the optimization problems using the interior-point algorithm implemented by the $\mathrm{fmincon}$ function in MATLAB software. Note that, since we already know that the optimal distribution is discrete with a finite number of mass points, the numerical optimization is over the position, the probabilities and the number of the mass points. Hence, there are $2m$ parameters to be optimized, where $m$ denotes the number of the mass points. Given the constraints, the algorithm is fed with an input of two mass points, i.e., $m=2$ and a random guess for the respective probabilities and positions of the mass points. Once the algorithm outputs the local optimal solution for the positions and their respective probabilities,\footnote{Note that despite the fact that the problem is concave with respect to probability laws, however,  for a given number of mass points $m$, the problem is not concave and the obtained solution is not guaranteed to be a global one.} the answer is validated by checking the constraints and the corresponding necessary and sufficient KKT conditions in (\ref{E70}) or (\ref{E7}). If the conditions are not satisfied, the initial guess is changed. We continue changing the initial guess for a large number of times. If the KKT conditions are not satisfied yet, the number of mass points is increased by one. The algorithm runs until such an input is found.

\begin{figure}
\begin{centering}
\includegraphics[scale=0.26]{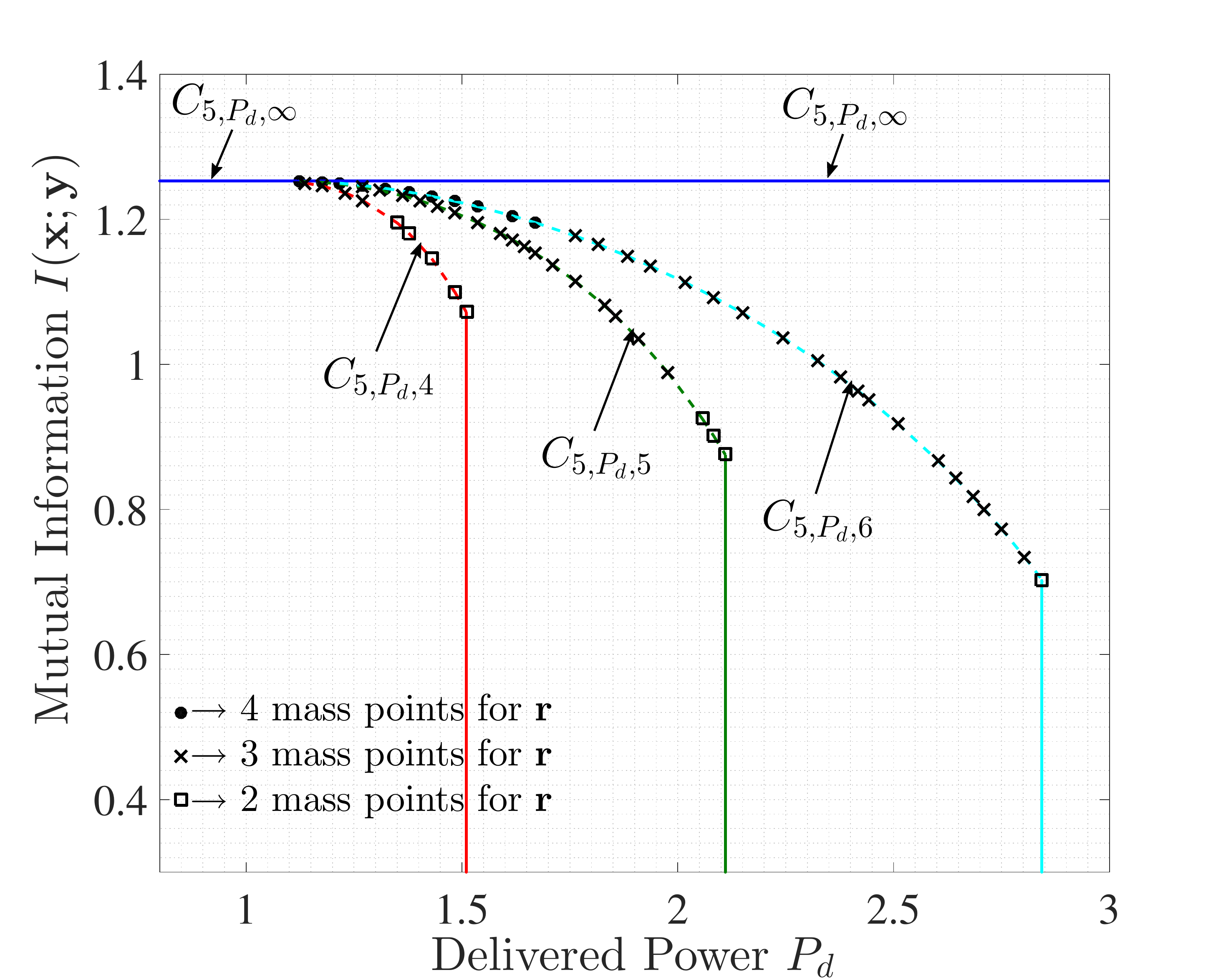}
\caption{Mutual information $I(\pmb{x};\pmb{y})$ corresponding to the optimal solutions of (\ref{E3}) with respect to different values of the RDP constraint $P_d$  with PA constraints $r_p=4,~5,~6$ and $r_p=\infty$ and AP constraint $P_a=5$.} \label{F1}
\par\end{centering}
\vspace{0mm}
\end{figure}

\textit{Illustration of the numerical results}:  In Fig. \ref{F1}, simulation results for the problem (\ref{E3}) with $P_a=5$ and $g(r)=0.01(r^4+r^2+1)$ are illustrated. The horizontal solid line related to $C_{5,P_d,\infty}$ corresponds to the AWGN channel capacity under an AP constraint $P_a=5$ achieved by only a CSCG distribution. The horizontal dashed line related to $C_{5,P_d,\infty}$ corresponds to the capacity under an AP constraint $P_a=5$, which is not achievable, however, can be approached arbitrarily (see Theorem \ref{T0}). $C_{5,P_d,4}$, $C_{5,P_d,5}$ and $C_{5,P_d,6}$ correspond to the optimal solution in (\ref{E3}) for $r_p=4,~5$ and $6$, respectively. It is observed that by increasing the PA constraint $r_p$, the capacity tends to the one corresponding to $r_p=\infty$. This observation is inline with Remark \ref{Rem1}, that increasing $r_p$, reduces the dependency of the capacity on $r_p$. Note that given the value of $r_p$, the amount of maximum RDP at the receiver is limited. This is the reason for the vertical lines corresponding to $C_{5,P_d,4}$, $C_{5,P_d,5}$ and $C_{5,P_d,6}$. It is observed that the number of optimal mass points $m$ for $\pmb{r}$ is decreased with $P_d$.

\begin{figure}
\begin{centering}
\includegraphics[scale=0.26]{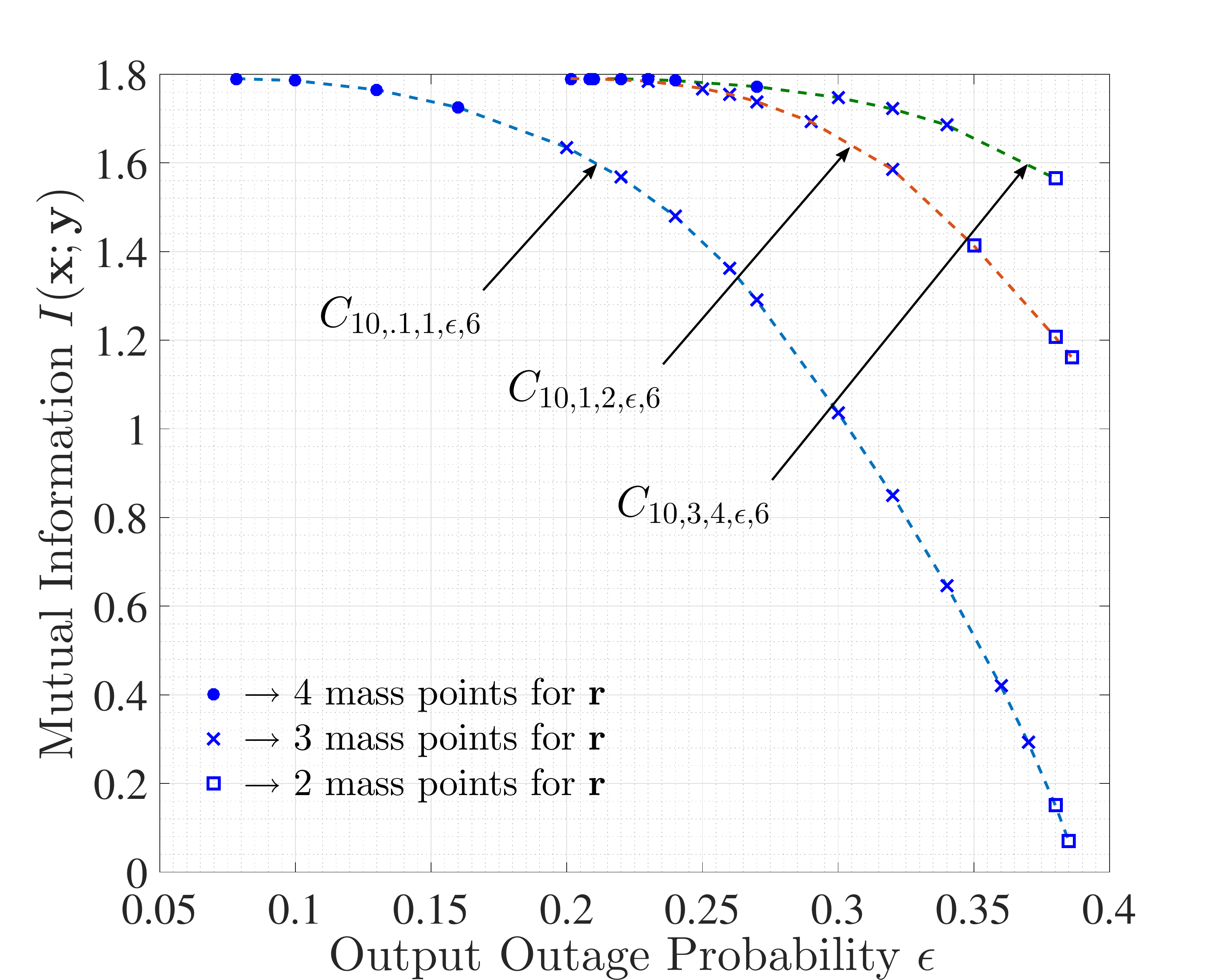}
\caption{Mutual information $I(\pmb{x};\pmb{y})$ corresponding to the optimal solutions of (\ref{E1}) with respect to different values of the OOP constraint $\epsilon$  with PA constraint $r_p=6$, AP constraint $P_a=10$.} \label{F2}
\par\end{centering}
\vspace{0mm}
\end{figure}

In Fig. \ref{F2}, similar results to the problem (\ref{E1}) are obtained with AP $P_a=10$, PA $r_p=6$ for different values of $A_l,A_u$ and $\epsilon$. It is observed that by increasing $\epsilon$ (equivalently decreasing the outage probability), the rate is decreased. It is also observed that the number of mass points decreases with $\epsilon$.

We note that, the algorithm used for finding the optimal distributions is extremely sensitive to the first guess as the number of mass points $m$ increases. This is due to the fact that optimization of the capacity given that the number of mass points $m$ is fixed, is not a concave function. This accordingly, makes the problem computationally demanding with $m$.

\section{Conclusions}\label{Sec_Conclusion}
In this paper, we studied the capacity of a complex AWGN channel in the presence of a nonlinear energy harvester at the receiver. Motivated by the two nonlinear models of the energy harvester introduced in the literature, two sets of constraints were studied. First, we considered the capacity problem under AP, PA and RDP constraints. We showed that the capacity under the AP and RDP constraint, is the same as the capacity under AP constraint that can be either achieved or approached arbitrarily. In line with the similar results in the literature, we showed that including the PA constraint causes the amplitude of the optimal input to be discrete with a finite number of mass points. Next, we considered the capacity problem under AP, PA and OOP constraints, where similar results were derived.
\appendix

\bibliographystyle{ieeetran}
\bibliography{ref}

\end{document}